# Characterization of one-dimensional quantum channels in InAs/AlSb

C. H. Yang[1], M. J. Yang[2], K. A. Cheng[1], and J. C. Culbertson[2]

[1]Dept. of Electrical and Computer Engineering, Univ. of Maryland, College Park, MD 20742

[2]Naval Research Laboratory, Washington, DC 20375

*Abstract*

We report the magnetoresistance characteristics of one-dimensional electrons confined in a single InAs quantum well sandwiched between AlSb barriers. As a result of a novel nanofabrication scheme that utilizes a 3nm-shallow wet chemical etching to define the electrostatic lateral confinement, the system is found to possess three important properties: specular boundary scattering, a strong lateral confinement potential, and a conducting channel width that is approximately the lithography width. Ballistic transport phenomena, including the quenching of the Hall resistance, the last Hall plateau, and a strong negative bend resistance, are observed at 4K in cross junctions with sharp corners. In a ring geometry, we have observed Aharonov-Bohm interference that exhibits characteristics different from those of the GaAs counterpart due to the ballistic nature of electron transport and the narrowness of the conducting channel width.

## 1. Introduction

Electrons confined in low dimensional systems have been under intensive investigation both theoretically and experimentally in recent years.[1,,3] One popular sample system involves semiconductor heterojunctions, particularly the GaAs/Al$_x$Ga$_{1-x}$As high electron mobility transistor structure grown by molecular beam epitaxy (MBE). The two dimensional (2D) electrons at the heterojunction interface exhibit high mobility, long mean free path ($l_e$), and long phase coherence length ($l_f$). A typical fabrication method to confine laterally electrons into quasi-one and -zero dimensions is the use of split-gates:[4,5] Schottky metal gates deposited at the surface can deplete electrons below, thereby defining 2D electrons into an arbitrary pattern. Nanostructures fabricated by this approach have revealed interesting new quantum phenomena[1]. However, in the Schottky split gate approach, due to the material properties and confinement electrostatics, the confinement potential is parabolic[6] in the lateral directions. The resulting 1D level spacing is at most a few meV. In addition, it is difficult to control the channel width when it is below 100 nm in this materials system. New techniques for patterning narrower (< 100 nm) quantum wires and dots are therefore of fundamental interest.

Recently, we have reported a new technique[7] for nanofabrication in the InAs/GaSb/AlSb 6.1Å material system[8] that utilizes the large difference in the surface Fermi level pinning positions for InAs ($E_f^s(InAs)$) compared with that for AlSb. While this fabrication technique is of interest in it own right, it is important that electrons in the resulting quantum wires maintain long elastic mean free paths and phase coherence lengths so that they provide a useful system for experimental studies and for device applications. We report in this paper a systematic study of transport characteristics as a function of the width of quantum wires fabricated using our newly developed technique, which employs high resolution electron beam lithography and wet chemical etching in high mobility AlSb/InAs/AlSb single quantum wells (QWs).

We demonstrate that this system possesses three important properties that are crucial for practical applications in building a complex coherent circuit within a mean free path. First, the lateral boundary scattering is found to be more than 95% specular, which means that InAs nanowires can maintain a long $l_e$. Second, as a result of the tight lateral 1D confinement and a smaller electron mass in InAs, the 1D sublevel spacing is found to be a few times bigger than in the GaAs system. Third, the conducting channel width ($W_c$) is found to be approximately only 30 nm narrower than the wire width ($W$). This makes it possible to tightly control the wire width, and most importantly, to place multiple nano devices, such as dots and rings in close proximity and within the coherence length for quantum interference experiments.

## 2. Nanofabrication of 1D wires

We will first describe the nanofabrication scheme here. The samples, grown by MBE on semi-insulating GaAs substrates, consist of a 100 nm AlSb bottom barrier, a 17 nm InAs QW, a 25 nm AlSb top barrier, and a three-layer cap (3 nm InAs / 3nm AlSb / 3 nm InAs). The capping layers are intentionally p-doped to $10^{19}$ cm$^{-3}$. As a result of a relatively low $E_f^s(InAs)$ and the doping profile, there are no free carriers in the 17 nm InAs single quantum well so the quantum well is insulating as-grown at 4K. Figure 1 (a) shows the self-consistent band bending for as-grown samples, where the lowest subband in the InAs QW is found to be ~100meV above the Fermi level. Simply by selectively removing the first thin p-doped InAs cap layer with a wet etch, the surface Fermi level becomes pinned on AlSb and shifts upward by half an electron volt. This results in a drastic change in the band bending as shown in Fig. 1 (b), and creates a conducting electron channel in the buried InAs quantum well. Quantum Hall plateaus and Shubnikov de Haas (SdH) oscillations, both characteristics of 2D electrons, are clearly observed on photo-lithographically patterned Hall bars. We obtained a 2D electron concentration ($n_{2D}$) of $4.92 \times 10^{11}$ cm$^{-3}$ and a mobility ($\mu_{2D}$) of $2.0 \times 10^5$ cm$^2$/Vs. The corresponding Fermi wavelength ($\lambda_F$), Fermi energy, and $l_e$, are calculated to be 36 nm, 43 meV, and 2.3 µm, respectively.[7] In the following, we will describe how we pattern 1D wires/rings by electron beam lithography.

Two sets of metallic alignment marks for coarse and fine alignments are first defined by photolithography. The sample is then coated with a layer of PMMA (Poly(methyl methacrylate)) resist, and exposed to a 40keV electron beam for device patterning. PMMA is a positive tone resist, i.e., the region exposed to electron beam will be removed after development. We then use PMMA as the etch mask, and the patterned opening is subjected to wet etching. A highly selective etchant removes the top 3nm InAs and stops at the AlSb cap. After the wet etching is accomplished, the PMMA is removed. Bonding pads are defined by the second electron beam lithography, where all the cap layers and the top AlSb barrier are etched off, followed by metal evaporation on the exposed InAs QW surface and lift-off. Figure 2 shows an AFM (atomic force microscopy) topographic image of a figure-8 structure, which has a diameter of 700 nm and $W = 80$ nm. The bright region is lower than the field by approximately 3 nm, consistent with the metallurgical thickness of the InAs cap. The grain feature in the field is the residual PMMA granules. A thorough dilution using acetone leaves the field with fewer PMMA particles. We find this fabrication technique convenient, as the pattern can be arbitrarily generated by computer graphics. The wet etching method is demonstrated to be highly selective. Most importantly, because we are only targeting a shallow, 3 nm etch depth, the uncertainly in the $W$ is limited to be within a few nanometers, as shown in Fig. 2. In the 17 nm InAs quantum well layer, the region underneath the field remains insulating at 4.2K, but that below the etched pattern contains electrons.

### 3. Magnetoresistance in Long Nanowires: Diffusive Transport

#### A. Specularity of Boundary Scattering

The lateral confinement of electrons underneath the etched area is electrostatic similar to the case of Schottky gate depletion approach. It is then expected that 1D electrons fabricated by our technique maintain a long $l_e$, and insensitive to $W$. In order to determine the specularity parameter, $p$, of the boundary scattering,[9] we have carried out magnetotransport measurements at $T = 4.2K$ using the Hall bar geometry. We fabricated eight Hall bars with different targeted wire width ($W_l$), including 5 μm, 1 μm, 700nm, 450nm, 300nm, 200nm, 100nm, and 80 nm. Each Hall bar has two channel lengths, $L = 15$ μm and $L = 4\ W_l$ (except for 80nm, $L = 5\ W_l$). The AFM inspection on the Hall bars shows $W » W_l + 30$nm, due to the proximity effect and the isotropic wet etching. All of the Hall bars are fabricated from the same wafer and are at most a few mm apart. Other than the built-in nonuniformity inherent in MBE growth, their structures are identical. The a.c. driving current is kept between 10nA to 100nA, roughly scaled with $W_l$, and the resulting voltage drop across two current leads is much less than the thermal energy. Fig. 3 shows the (a) longitudinal ($r_{xx}$) and (b) the transverse ($r_{xy}$) magnetoresistances for different $W_l$ with $L = 15$ μm. As discussed by Thornton *et al.*,[9] partially diffusive boundary scattering can result in a positive zero-field magnetoresistance, which reaches a maximum at $B_{max}$ that scales with the ratio of the cyclotron radius of Fermi electrons ($l_{cyc} = \dfrac{h}{qB\mathbf{1}_F}$) to the conducting width as $W_c \approx 0.55 \, (\dfrac{h}{qB_{max}l_F})$. On the other hand, junction scattering can also cause a distinct magnetoresistance peak[10-12] below $B_{crit}$, where $B_{crit}$ is the magnetic field beyond which the cyclotron orbit is smaller than the channel width, i.e., $2\, l_{cyc} < W_c$. $B_{crit}$ for different $W_l$, assuming $W_l = W_c$, are listed in Table 1, where $B_{max} = 0.275 B_{crit}$. The way to distinguish these two phenomena is by investigating wires with different lengths.[13] Magnetoresistance peaks resulting from diffusive boundary scattering are observable only when $L > l_e$, while that from junction scattering is more obvious in the ballistic regime, $L < l_e$. As shown in Fig. 3 (a), the data do not display a magnetoresistance peak except for a weak kink when $W_l \leq 700$nm, where the corresponding $B_{max}$ is indicated by the arrow. (By comparison, magnetoresistance peaks can be clearly identified for the short segments due to the cross junction scattering when $100\text{nm} \leq W_l \leq 1\mu\text{m}$. This will be discussed in section **4B**). As a result, the lower bound for the probability of specular scattering is determined to be $p = 95\%$ from $1-p = 700$ nm/15 µm. The observed nearly 100 percent specularity indicates that $l_e$ is predominately limited by impurity scattering in long, narrow wires.

The observed high specularity is expected, since with our new fabrication scheme no additional surface scattering centers are created and the edge of the 1D electrons is defined by an electrostatic potential. Consequently, the electron mean free path is expected to be insensitive to the wire width, and the material quality will not be degraded by the additional lateral confinement. This is further confirmed by examining the zero-field mobility. Figure 4 summarizes the density and zero-field mobility as a function of $W_l$ for $L = 15$ µm, where the density is obtained from the slope of $r_{xy}$ at B > 1T and minimum of $r_{xx}$ at B > 3T. For Hall bars with $W_l > 200$ nm, the density remains more or less constant. We ascribe the slight variation in the density to the nonuniformity of the MBE grown material. For $W_l < 200$ nm, we observed an apparent reduction in density, which is attributed to the combination of two effects: lateral quantization and the slightly narrower conducting channel width due to the lateral depletion width. The lateral depletion width is estimated to be 30nm, and this makes $W_c \gg W_l$. The zero-field resistivity does not show a strong temperature dependence for $W_l \geq 300$ nm at T = 4K. However, due to weak and strong localization effects,[14,15] $r_{xx}$ (B=0) increases by 8%, 43%, and 70%, respectively, for $W_l$ = 200 nm, 100nm and 80nm as temperature is decreased from 4K to 2K. This makes the determination of the zero-field mobility less meaningful for $W_l \leq 200$ nm, especially for $W_l$ = 80 nm. The inset of Fig. 4 depicts the dependence of the mobility on the density at 4K. We found that the mobility scales with the density in the same way as in 2D systems,[16] as indicated by the solid line in the inset. In other words, again, lateral confinement does not introduce significant diffusive boundary scattering that can localize the electrons, which is consistent with the observed high specularity of boundary scattering. As a result of the reduction of carrier density and mobility for narrow $W_l$, however, $l_e$ is reduced slightly (see Table 1.)

**B. Magneto-Depopulation of 1D Channels**

The fact that material quality does not deteriorate from the additional lateral confinement easily brings our InAs wires into the effectively pure 1D metal regime ($W_l \ll l_e$) and makes it possible to observe magnetic depopulation of 1D subbands. When a 1D system is subjected to a perpendicular magnetic field, the 1D density of states evolves into 2D Landau levels, as depicted in the inset of Fig. 5 (b) for a channel width of 100nm. A magnetic field imposes additional parabolic confinement on the electrons and results in hybrid electric-magnetic subbands. The hybrid subbands eventually evolve into Landau levels at high magnetic field when the strength of the magnetic parabolic confinement surpasses the (lateral) electrostatic one. The crossover occurs when the diameter of the cyclotron orbit for each Landau level becomes smaller than the channel width. This occurs when $2\sqrt{\frac{\hbar}{eB}(2N-1)} \leq W_c$, where the Landau level (or 1D channel) index $N = 1, 2, 3…$ Evidence for the 1D sublevel formation is provided by magneto depopulation experiments, where the magnetoresistance oscillations are no longer periodic in $B^{-1}$. Figure 5 shows the 1D channel index and the experimental $r_{xx}$ as a functional of $B^{-1}$. The oscillations in $r_{xx}$ are masked by the quantum interference, including the universal conductance fluctuations[17] (UCF) and the structural resonance effect[18] (which will be discussed in section **4A**). This makes the determination of minimum in $r_{xx}$ difficult at low fields. We have filtered out the fast components of $r_{xx}$ due to quantum fluctuations, and assigned the proper 1D channel index based on the larger oscillation amplitudes. Ten (five) 1D sublevels are observed for $W_l$ = 200nm (100nm), in good agreement with the estimates of $N$ = 10 (4) from $\frac{2W_l}{l_F}$. It is clear that the 1D index starts to deviate from the 2D linear relationship around 1.01 $tesla^{-1}$ at $N = 8$ for $W_l$ = 200 nm and 0.67 $tesla^{-1}$ at $N = 4$ for $W_l$ = 100 nm. The corresponding cyclotron diameters are calculated to be 200nm and 111nm, respectively, implying that the exact conducting width is very close to the targeted one, i.e., $W_c \approx W_l$.

Although the data presented in Fig. 5 are similar to what have been observed in GaAs nanowires, they cannot be fit by assuming a harmonic potential confinement. We follow the procedures laid out by Berggren et al.,[6] and impose two constraints in the fitting parameters. First, the calculated effective width $\widetilde{W}$ cannot exceed 20% of $W_l$. Second, the effective 2D density ($\frac{n_{1D}}{\widetilde{W}}$, where $n_{1D}$ is the 1D electron density) cannot exceed the bulk 2D value, $4.9\times10^{11}$ cm$^{-2}$. The solid curves in Fig. 5 are the best fits to the linear regime obtained by assuming an effective mass of $0.028m_0$. The harmonic quantization energy, $n_{1D}$, and $\widetilde{W}$ are obtained, respectively, to be 3.5 meV, $8.9\times10^6$ cm$^{-1}$, and 182 nm for $W_l$ = 200 nm, and 5.8 meV, $4.1\times10^6$ cm$^{-1}$, and 100 nm for $W_l$ = 100 nm. Although the parameters obtained are physical, the numerical fits to the high indices show a deviation. The fitting results indicate that the lateral potential shape is steeper than parabolic. The total lateral depletion width (including two boundaries) is empirically determined to be 30nm. An accurate description of the potential profile requires a 2D self-consistent calculation and an understanding of the detailed impurity distribution. Nevertheless, this fitting exercise does reveal, as expected, a 1D subband structure with a large quantization energy.

## 4. Magnetoresistance in Short Nanowires: Quasi-Ballistic and Ballistic Transport

### A. Quantum Interferences

As a result of the short lateral depletion length , nano devices can be fabricated close by one another. For example, the spacing of the voltage probing leads for $W_l$ = 100nm is as short as 400nm, and in principle can be further reduced. As the probe spacing approaches and eventually becomes shorter than $l_e$, the magnetoresistance exhibits different characteristics from what have been described in the previous section. Figure 6 (a) shows the longitudinal ($R_{12,34}$) resistance, where current is passed from lead 1 to 2 and the voltage difference is measured from 3 to 4. The lead numbers are indicated in the inset of Fig 6 (b) that shows an AFM image of a Hall bar. The arrows mark the corresponding $B_{crit}$ for different $W_l$. For 200 nm ≤ $W_l$ ≤ 1μm, $R_{12,34}$ exhibits two features: a magnetoresistance peak below $B_{crit}$, and a reproducible quantum interference signal that slightly distorts the SdH oscillations. The resistance peak is a result of junction scattering of ballistic electrons, which will be elaborated on in section **4B**. The interference signal manifests two signatures: aperiodic, fast variation UCF at low fields, and quasi-periodic slow fluctuations due to the structural resonance at high fields. However, when the width and length are further reduced, UCF are not discernable due to the nature of the ballistic transport. Instead, the interference due to the structural resonance dominates even at low fields ($W_l$ = 100nm) and eventually exceeds the SdH strength for whole range of magnetic field ($W_l$ = 80nm). The structural resonance at high field where only the lowest Landau level is occupied, have a spacing of ~0.11 tesla (~0.15 tesla) for $W_l$ = 100nm (80nm). It is interesting to note that the period of the Aharonov-Bohm (AB) interference corresponds to an enclosed area, which is approximately the "channel space," i.e., 100nm × 400nm (80nm × 400nm).

The electron phase coherence length is an important parameter that needs to be determined for our nano system. Electron interference phenomena such as weak localization, UCF, and AB oscillations in ring structures[19] have been used to determine $l_f$. However, the suppression of weak localization by a magnetic field is masked by UCF for 200 nm ≤ $W_l$ ≤ 1μm in both short and long segments. Consequently, we estimate $l_f$ from the root-mean-square of the resistance fluctuation here and from AB oscillations (discussed in the next section.) It is expected that UCF would disappear as $L$ is reduced and approaches $l_e$. When $L \gg l_e$, the strength of UCF depends on the relative length scale of $L$ and $l_f$, as expressed by[20]

$$\Delta G \approx \frac{\Delta R}{R^2} = \sqrt{12} \frac{e^2}{h} \left(\frac{l_f}{L}\right)^{3/2} \left[1 + \frac{9}{2p}\left(\frac{l_f}{l_T}\right)^2\right]^{-1/2}.$$

Here the thermal length is defined as $l_T = (\hbar D / k_B T)^{1/2}$ and $D = v_f l_e$, where $D$ and $v_f$ are the 1D diffusion constant and the Fermi velocity, respectively. For $W_l =1\mu m$ (700nm), $\Delta R/R^2$ is $6\times10^{-5}$ ($7\times10^{-5}$), corresponding to 1.5 (1.8) $e^2/h$. The electron phase coherent length is estimated to be 2.5μm (2.0μm) at 4K for our 1D wires. It has been shown that at 4K, $l_F$ is dependent on temperature as $T^{-1}$.[21] It is then expected that in InAs quasi 1D system, $l_F$ will be longer than 10 μm at sub Kelvin temperatures, comparable to what has been measured in GaAs/AlGaAs 1D systems.

It is worth mentioning that the sheet resistivity obtained from the short segment is always larger than that from the long one by, e.g., 10% for $W_l =1\mu m$, and increases monotonically to 70% for $W_l = 200$nm. This observation is accounted for by non-local effects.[22] When $L$ approaches or becomes less than $l_F$, both the voltage and current leads become an integral part of the device under inspection, which makes the effective size of device larger than what been defined by lithography. In addition, when $L < l_e < l_F$, the resistance in the short segments decreases as the temperature is decreased, opposite from what has been observed in the long segments. For example, the resistance decreases by 13%, 27%, and 37%, respectively, for $W_l = 200$ nm, 100nm and 80nm as temperature is decreased from 4K to 2K. This is explained by noting that the system approaches ballistic transport at a lower temperature.

### B. Quenching of the Hall Effect and Bend Resistance

Experimental verification of ballistic transport phenomena in InAs nanowires is discussed in the following. In cross junction devices, we have observed the quenching of the Hall effect, a weak "last plateau" feature,[10-12] and negative band resistance;[23] all of these indicate that the system is in the ballistic transport regime and that there is a substantial collimation effect. The Hall resistances ($R_{12,46}$) for 200 nm, 100 nm, and 80nm wide cross junctions are shown in Fig. 6 (b). The observed low field Hall resistance can be negative, zero, or super-linear, and is strongly dependent on the exact, detailed geometry at the junction. Such a deviation from ideal classical Hall characteristics at low fields is observed only in the two narrowest wires, $W_l$ = 100 nm and 80 nm, reflecting that the junction corners are relatively sharp.[12,24] In other words, sharp corners suppress quenching for $W_l \geq$ 200 nm. As the magnetic field increases, the last plateau feature in $R_{12,46}$ is accompanied by a magnetoresistance peak in $R_{12,34}$. Both result from electron guiding due to the Lorentz force. Because of the sharp corners at the cross junction, the last plateau feature is weak in our samples. It disappears at high fields when $W_l > 2\ l_{cyc}$ as the system is brought to the 2D regime.

In our narrow cross junctions, the collimation effect also generates a strong negative bend resistance, as shown in Fig. 7 for $W_l$ = 100 nm. Based on the classical picture, $R_{16,42}(B=0)$ should be close to zero. However, in the ballistic regime, electrons injected from lead 6 more efficiently transmit to the modes in lead 4 than to those in lead 1, and this results in a negative voltage of $V_{42}$. Under a perpendicular magnetic field, the Lorentz force changes the trajectory of electrons and eventually destroys the negative bend resistance. The field that destroys the negative bend resistance also marks the onset of the last Hall plateau feature. We find the same critical fields taken from the same junction at ~0.5 tesla. The observed critical field is larger than what has been observed in GaAs/AlGaAs nano wires, and is in good agreement with a calculation assuming a hard-wall confinement and sharp corners.[12] The four-terminal resistance $R_{16,42}$ recovers to 2D SdH-like oscillations at high field, as expected for a Van der Pauw geometry. We note that while the bend resistance returns to the 2D characteristics under a sweeping magnetic field, there is no significant overshoot, again indicating that our junction is sharp.

## 5. Magnetoresistance Characterization in Rings: Aharonov-Bohm Oscillations

The AB interference effects have been studied extensively in mesoscopic ($l_F$, $l_e \ll L \ll L_f$) metallic[19] and semiconducting[25] rings. In narrow metal wires, electrons show no size quantization effects even at the lowest temperatures. They suffer from elastic scattering (with a typical $l_e \sim$ 1 to 10 nm) and traverse the ring diffusively. Yet, AB interference experiments make it evident that there is long range phase coherenece. Experiments on semiconductor rings, particularly those fabricated from high mobility GaAs-heterojunctions, have gained insights into the mechanism of the AB interference. In combination with our new nanofabrication scheme and a resulting hard-wall confining potential, the diameter of the InAs rings can be made smaller than $l_e$ and with much better control and knowledge of the wire width than for the GaAs counterpart. As the system enters the ballistic transport regime, the AB oscillations exhibit characteristics different from what have been commonly observed in GaAs rings.

Figure 8 (a) shows the magnetoresistance of a 500nm ring with $W_l = 100$ nm at 2K, where two traces were recorded simultaneously using voltage leads on the opposite sides of the current channel. Aside from the difference in the absolute resistance value due to non-local effects, these two traces consist of AB oscillations superimposed on an asymmetric negative magnetoresistance with step-like features, e.g., at ± 250 mtesla. The system returns to 2D behavior above 2 tesla and exhibits SdH oscillations similar to the middle curve in Fig. 7 (a). The large negative magnetoresistance around zero field is attributed to the reduction of backscattering at the inner wall of the loop, and the step-like features are accounted for by cyclotron trapping.[26]

When the large negative background resistance is subtracted, the oscillations corresponding to $h/e$ and $h/2e$ fluxes are clearly observed in Fig. 8 (b). The strong oscillations occur up to one tesla or so, and remain discernable until ~2 tesla, consistent with $B_{crit}$ estimated for $W_l = 100$ nm in Table 1. The Fourier spectrum of oscillations in the range of ± 250 mtesla, shown in the inset of Fig. 8 (a), exhibits four harmonic components. The fundamental frequency is centered at 48 tesla$^{-1}$, implying an average diameter of 503 nm. It is found that the amplitude of the fundamental peak, $h/e$, is twice the second peak, ten times the third peak and about twenty times the fourth peak. Unlike what has been observed in GaAs rings, the amplitudes of these four peaks do not follow an exponential dependence expected for the diffusive transport. This is attributed to the fact that the fundamental and the second harmonic peaks result from ballistic and quasi-ballistic electrons, respectively. Consequently, we have used the 3$^{rd}$ and the 4$^{th}$ harmonics to fit an exponential decay formula, $\exp(-L/L_f)$, where $L = i \times \pi d_{ring}$, $i$ is the order of the harmonics and $d_{ring}$ is the diameter of the ring. We obtain $L_f = 3.7$ μm at 2K, compared with $L_f = 2.5$ μm obtained at 4.2K for the same $W_l$ as discussed in section **4A.**

In the case of diffusive metallic rings where $W = W_c$, the width of the $h/e$ fundamental peak is found to be determined by the inner and outer diameters of the conducting wire, i.e. $\Delta B^{-1} \approx \frac{p\, d_{ring} W_c}{h/e}$. As a result, it is a common practice to estimate $W_c$ from the width of the $h/e$ peak for GaAs rings, where the knowledge of $W_c$ is lacking due to the fact that $W \gg W_c$. The full width of the $h/e$ peak in our InAs ballistic ring is 13.6 tesla$^{-1}$, from which we can infer $W_c = 36$ nm as opposed to 100nm which was obtained from the previous discussion in section **3B**. If the conducting channel were indeed 36 nm, we should have only one 1D sublevel occupied, in contractiction to the unambiguous observation of filling factor 6 in $R_{12,34}$ at 2.25 tesla. A similar experiment done on a one-micron InAs ring with $W_l$ = 100 nm also shows a much narrower width of the *h/e* peak than expected, 20 tesla$^{-1}$ as opposed to 76 tesla$^{-1}$. One possible explanation of the observed narrow width of the Fourier power peaks is that the AB quantum interference comes mainly from the lowest 1D subband, whose electron wavefunction is concentrated in the middle of the channel.

Another characteristic, which is different from that in GaAs rings, is that the position of the fundamental peak is insensitive to the magnetic field. We have carried out Fourier transformation on the magnetoresistance with 0.5 tesla intervals up to 2 and 2.5 tesla for 500nm and 1 µm diameter rings, respectively. The power amplitude of the h/e peak decreases as the magnetic field is increased, but the peak position remains approximately the same. This observation can be accounted for by the lateral confinement potential.

## 6. Conclusions

We have performed systematic and detailed magnetoresistance measurements on InAs nanometer wires and rings fabricated by a novel scheme that employs a p-doped cap layer to create an insulating sample to start with. Based on studies with various wire widths down to 80 nm and lengths down to 400 nm, we found that the lateral confinement leads to predominately specular boundary scattering, and as a result the elastic mean free path is not substantially degraded when the system is patterned into nanometer-wide quantum wires. In addition, the conducting channel width is approxmately 60nm narrower than the pattern width defined by wet-etching. A number of ballistic transport phenomena are observed in the Hall bar geometry at 4K, including the quenching of the Hall resistance, the last Hall plateau, and a negative bend resistance. Their dependence on the magnetic field indicates a relatively sharp corner in the cross junction. Quantum phenomena such as magnetic depopulation of 1D sublevels and universal conductance fluctuations are also observed in InAs wires.

We have also studied the AB quantum interferences in the ballistic regime in which the diameter of rings is less than the elastic mean free path and also the phase coherence length. The AB oscillations in our ballistic InAs rings exhibit characteristics different from those reported in GaAs rings. First, due to the smallness of the ring size, we have observed up to the 4$^{th}$ harmonic component, *h/4e*, in the Fourier power spectrum of the AB oscillations. However, as a result of the ballistic nature of the electron transport in the fundamental and the second harmonic components, the amplitudes of the four harmonic peaks do not follow a simple exponential decay as expected by the diffusive picture. Second, unlike in the case of diffusive rings, the full width of the *h/e* peak in our ballistic InAs rings is much narrower than that estimated from the channel width. Third, the position of the *h/e* peak does not shift toward lower frequencies as the magnetic field is increased.

In summary, we have demonstrated a nanofabrication scheme leading to a 1D system that possesses desired properties for nanodevices, including specular boundary scattering, a strong lateral confinement potential, and a conducting channel width that is approximately 30nm narrower than the wire width. This InAs nano system is able to provide additional insights into ballistic transport.

## 7. Acknowledgement

The authors thank Dr. T. L. Reinecke for critical reading of the manuscript and Dr. Y. B. Lyanda-Geller for fruitful discussion. This work has been supported in part by ONR/NNI and LPS/NSA.

**Table 1.** Sample parameters for different targeted channel width $W_l$. The critical magnetic field $B_{crit}$ is calculated by $W_l = 2\, l_{cyc} = 2(h/e\, l_F\, B_{crit})$.

| $W_l$ (nm) | 5000 | 1000 | 700 | 450 | 300 | 200 | 100 | 80 |
|---|---|---|---|---|---|---|---|---|
| $n_{2D}$ ($10^{11}$ cm$^{-2}$) | 4.92 | 4.71 | 4.19 | 4.29 | 4.01 | 4.04 | 3.31 | 2.7 |
| $m_{2D}$ ($10^3$ cm$^2$/Vs) | 200 | 186 | 149 | 143 | 139 | 139 | 143 | — |
| $l_F$ (nm) | 36 | 37 | 39 | 38 | 39 | 39 | 44 | 48 |
| $l_e$ (μm) | 2.3 | 2.1 | 1.6 | 1.5 | 1.5 | 1.5 | 1.4 | — |
| $B_{crit}$ (tesla) | 0.046 | 0.23 | 0.30 | 0.48 | 0.70 | 1.05 | 1.90 | 2.14 |

**Figure Captions**

Fig. 1 Self-consistent band bending for (a) as-grown and (b) etched InAs/AlSb quantum wells, where energy is referred to the Fermi energy. The two solid lines in the InAs quantum well and in the 3nm InAs capping layers indicate the first two subbands.

Fig. 2 AFM image of a figure-8 geometry with a diameter of 700nm and $W$ = 80nm. The bright region is lower than the field by approximately 3nm.

Fig. 3 (a) Longitudinal and (b) transverse magnetoresistance obtained from Hall bars with a channel length of 15 μm for various $W_l$. The arrows indicate the position of $B_{max}$ as discussed in the text. Note $B_{max}$ = 0.275 $B_{crit}$, where $B_{crit}$ for different $W_l$ are listed in Table 1.

Fig. 4 The effective 2D density and zero-field mobility as a function of $W_l$. The inset is the logarithmatic plot where the solid line indicates the power law $m_{2D} \propto n_{2D}^{1.5}$ dependence.

Fig. 5 The 1D channel indices obtained from the minima of oscillations as a function of $B^{-1}$ for $W_l$ = (a) 200 nm and (b) 100nm. The dashed linear lines describe the expected dependence of a 2D electron system, and the solid curves are the calculated results assuming a parabolic lateral confinement potential. The dashed curves in the inset are the calculated Landau levels including the nonparabolic effect of the InAs conduction band. The solid curves in the inset depict the evolution of the 1D sublevels to Landau levels for $W_l$ = 100nm.

Fig. 6 (a) Longitudinal $R_{12,34}$ and (b) transverse $R_{12,46}$ magnetoresistance for the short segment of the Hall bar, where the arrows indicate the $B_{crit}$ position. The inset is an AFM image of a 300nm Hall bar, where lead numbers are indicated.

Fig. 7 Bend resistance for a cross junction with $W_l = 100$nm. Note the asymmetric feature around $B = 0$ and SdH-like oscillations above 3 tesla.

Fig. 8 (a) The magnetoresistance of an InAs ring of a diameter of 500nm and $W_l = 100$nm. Two traces are taken simultaneously using two symmetric pairs of the voltage leads. The left inset shows the AFM image of a device on the same wafer. (b) The AB oscillations of $R_{12,34}$, where the smooth background resistance is subtracted. The Fourier power spectrum of the $\Delta R$ is displayed in the right inset of (a). The short marks on the top of $h/e$ and $h/2e$ peaks indicate the full width of the peak, while the long ones show the expected width based on $\mathbf{D}B^{-1} \approx \dfrac{\mathbf{p}d_{ring}W_l}{h/e}$ and $\dfrac{\mathbf{p}d_{ring}W_l}{h/2e}$, respectively.

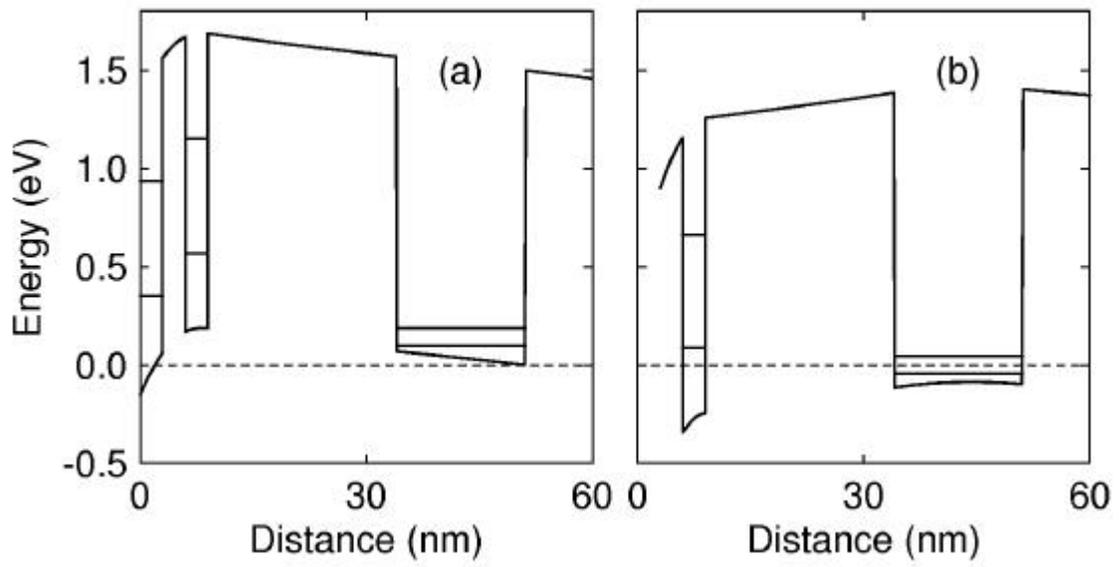

Figure 1 of 8

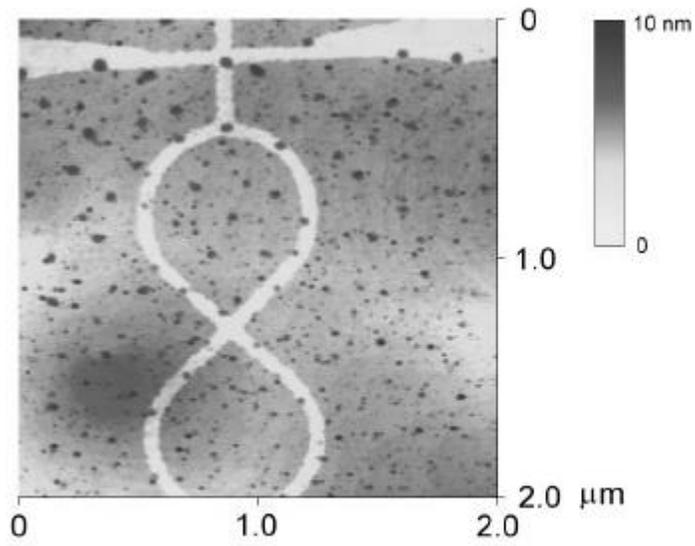

Figure 2 of 8

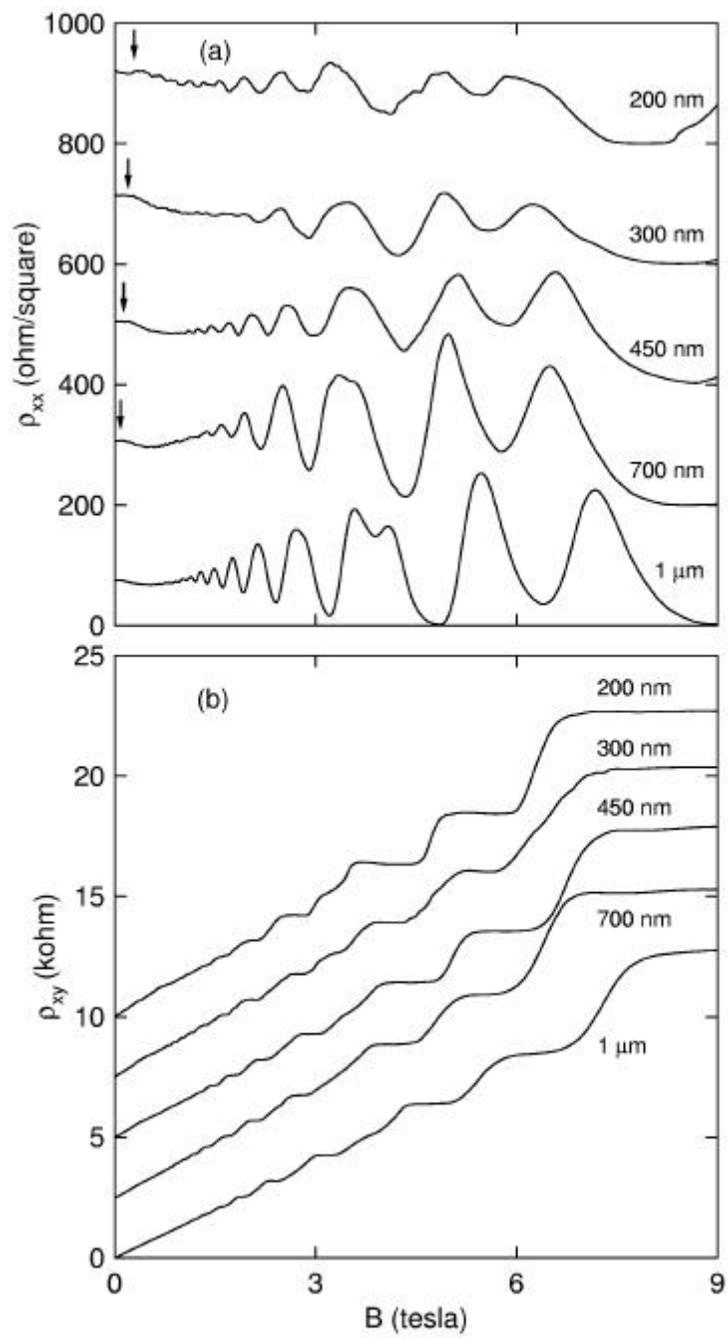

Figure 3 of 8

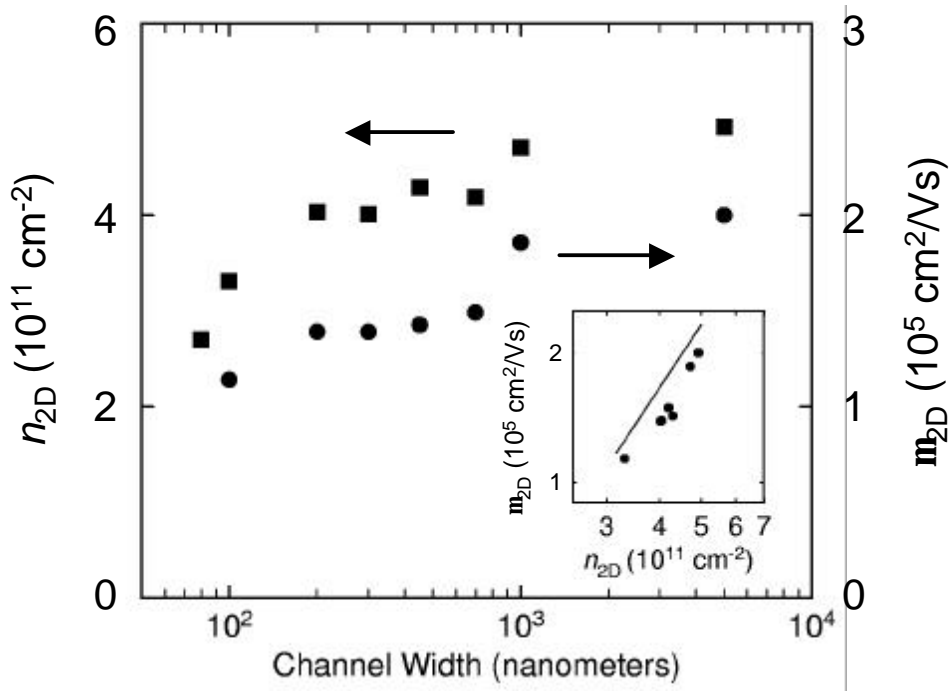

Figure 4 of 8

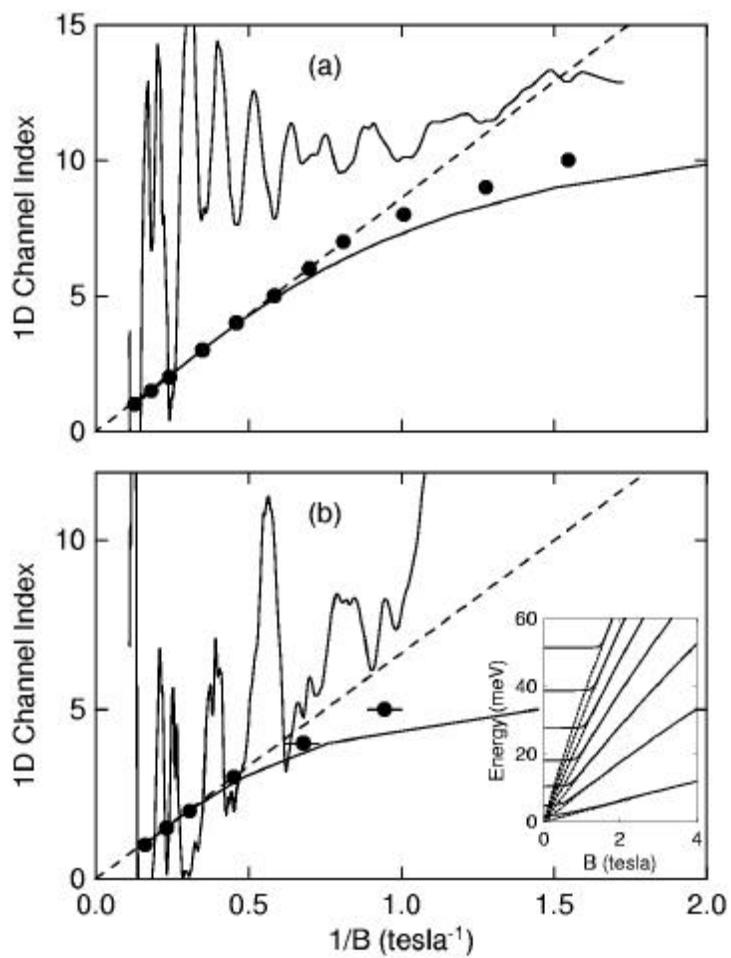

Figure 5 of 8

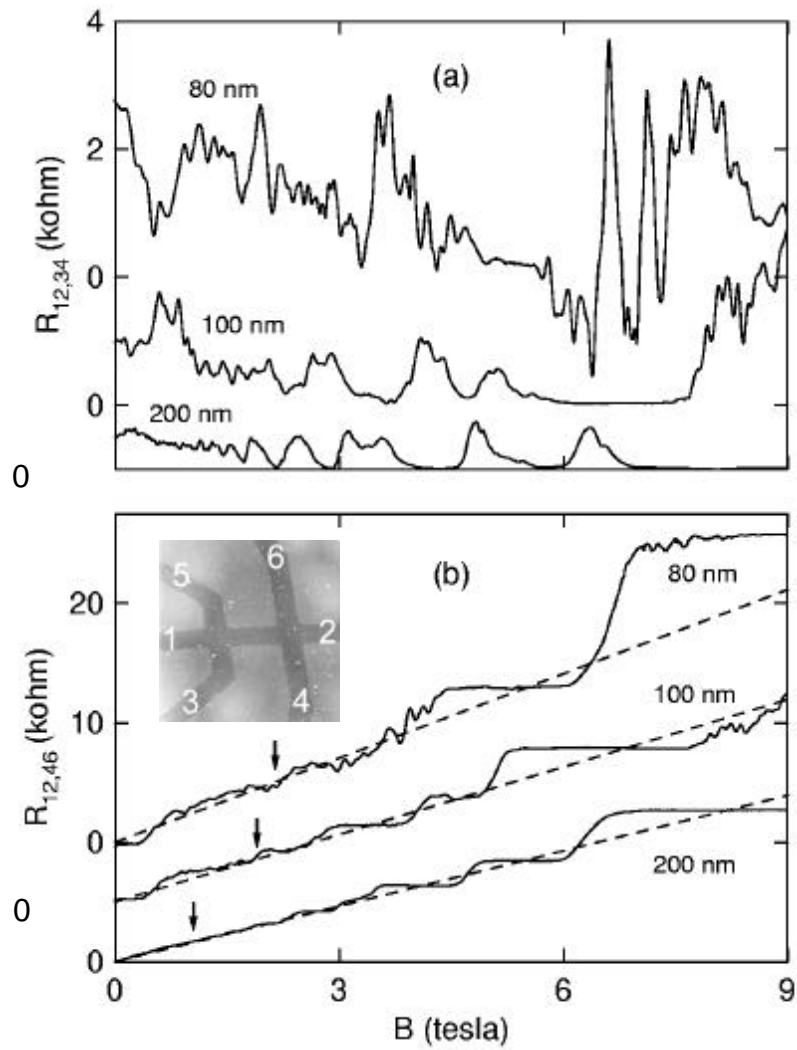

Figure 6 of 8

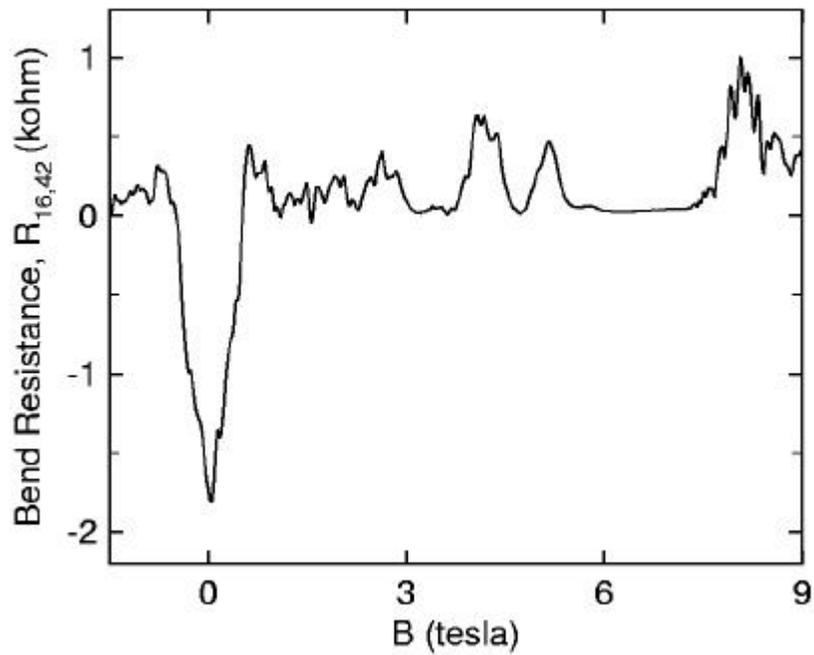

Figure 7 of 8

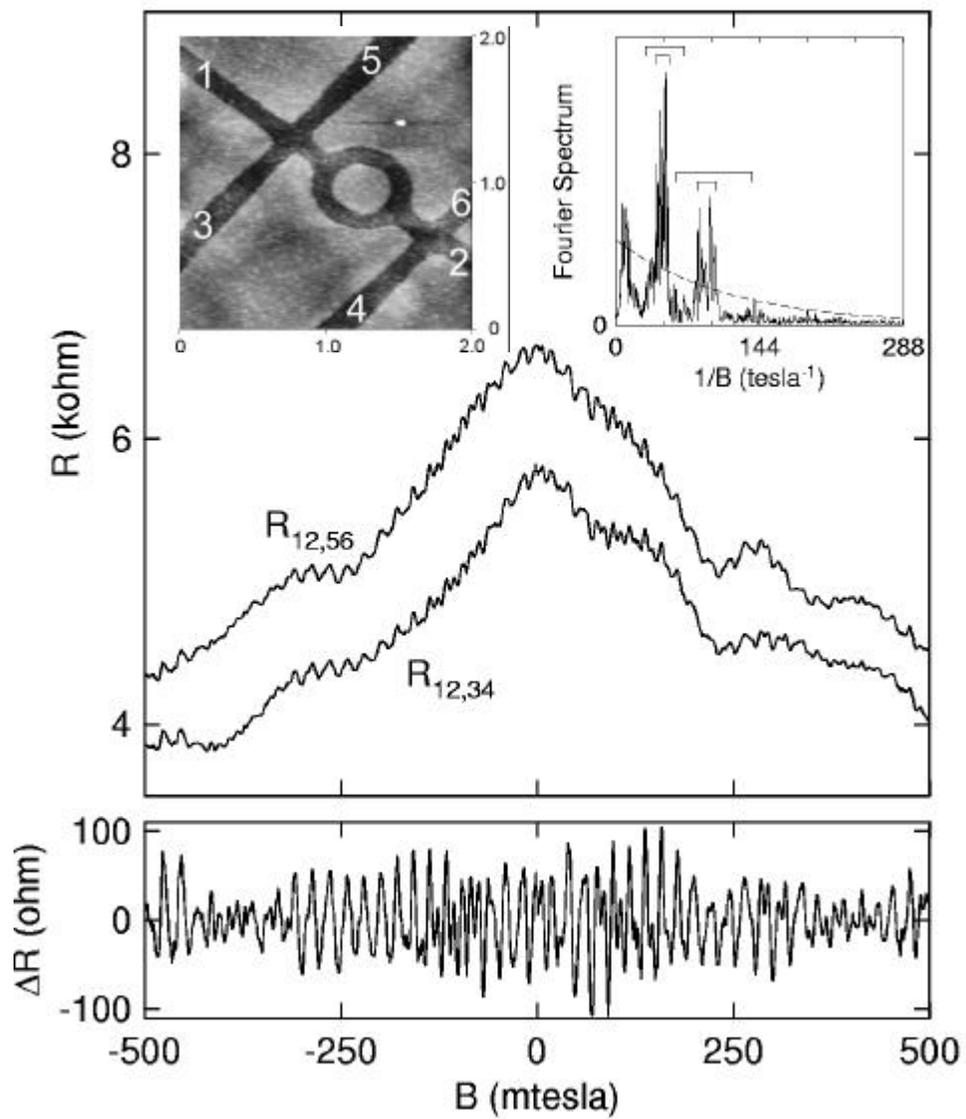

Figure 8 of 8